# Direct laser planting of hybrid Au-Ag/C nanostructures - nanoparticles, flakes and flowers

Alina Manshina[1]*, Anastasia Povolotckaia[2], Muhammad Bashouti[3], Alexey Povolotskiy[1], Yuriy Petrov[4], Igor Koshevoy[1,5], Silke Christiansen[3,6], Sergey Tunik[1], Gerd Leuchs[3,7,8]

[1] Institute of Chemistry, St. Petersburg State University, Universitetskii pr. 26, St. Petersburg, 198504, Russia
[2] Center for optical and laser materials research, Research park, St. Petersburg State University, Ulianovskaya St. 5, St. Petersburg, 198504, Russia
[3] Max Planck Institute for the Science of Light, Günther Scharowsky-Str. 1 Bau 24, D - 91058 Erlangen, Germany
[4] Interdisciplinary Resource Center for Nanotechnology, Research park, St. Petersburg State University, Ulianovskaya St. 1, St. Petersburg, 198504, Russia
[5] University of Eastern Finland, Joensuu, 80101, Finland
[6] Helmholtz Center Berlin for Materials and Energy, Hahn-Meitner-Platz 1, 14109 Berlin, Germany
[7] Department of Physics, Friedrich-Alexander-Universität Erlangen-Nürnberg, D-91058 Erlangen, Germany
[8] Department of Physics, University of Ottawa, 140 Louis Pasteur (130),Ottawa, ON, K1N 6N5, Canada

**ABSTRACT**
We demonstrate a new approach for forming hybrid metal/carbonaceous nanostructures in a controlled direct laser planting process. Au-Ag nanoclusters in amorphous or crystalline carbonaceous matrices are formed with different morphology: nanoparticles, nanoflakes, and nanoflowers. In contrast to other generation techniques our approach is simple, involving only a single laser-induced process transforming supramolecular complexes dissolved in solvent such as acetone, acetophenone, or dichloroethane into hybrid nanostructures in the laser-affected area of the substrate. The morphology of the hybrid nanostructures can be steered by controlling the deposition parameters, the composition of the liquid phase and the type of substrate, amorphous or crystalline. The carbonaceous phase of the hybrid nanostructures consists of hydrogenated amorphous carbon in the case of nanoparticles and of crystalline orthorhombic graphite of nanoscale thickness in the case of flakes and flowers. To the best of our knowledge this is the first demonstration of the fabrication of orthorhombic graphite with metal nano inclusions. The remarkable quality and regularity of the micron-sized nanoscale thickness single crystal flakes allows for cutting high resolution nano scale structures, which in combination with the metallic nano inclusions offer much design freedom for creating novel devices for nano photonic applications. The encouraging properties of the nanomaterials with different composition, size and shape stimulate the development of efficient synthesis strategies aimed at fine-tuning the functionality.

The combination of two or more nanoscale materials into a single structure or pattern may lead to new hybrid nanomaterial with enlarged functionality, determined not only by the additive combination of the initial properties of the components but often accompanied by a synergetic effect resulting in novel behaviors. For example, the association of luminescent nanoparticles (such as semiconductor quantum dots or rare-earth doped nanocrystals) with metal nanoparticles having plasmonic properties increases the luminescence intensity due to local electromagnetic field enhancement by the metallic nanocomponent [1]. Bimetal hybrid nanostructures possess improved catalytic performance in comparison with monometallic nanoparticles [2–5]. In addition, an even higher impact was found for ternary metal nanoclusters [6, 7]. A large effect is also demonstrated for metal/dielectric and metal/semiconductor nanohybrids, where special attention resides on metal/carbon hybrids due to the rich variety of carbon allotropes and morphologies [8–10]. The shape and morphology of the hybrid nanomaterials is important for the functional properties. Strong correlation of hybrid nanomaterials' morphology with their chemical, physical, electronic, optical, magnetic and catalytic properties opens ample opportunities from the viewpoint of diverse application areas. The special feature of such morphologically-manifold structures is a shape-selective dependence of their functionality. The other peculiarity of morphologically-complex nanostructures is their surface roughness, surface-to-volume ratio, high index facets, and electromagnetic field effects [11, 12]. By now, branched and 'flower'-shaped species of various organic and inorganic compounds, have been studied as active components for surface-enhanced Raman scattering (SERS), catalysis, possible carriers of drugs for medical applications, etc. [13–15]. Fine-tuning of functionality can be achieved by varying several key parameters: material composition, size and shape of the component parts and of composite hybrid species. All of these parameters determine the surface atomic arrangement and coordination permitting the control over activity, selectivity, and stability of hybrid nanomaterials. Currently, there are countless methods to create complex hybrid nanostructures from nanoparticles to flakes and flowers. But each method aims at creating specifically defined nanostructures; general approaches virtually do not exist at the moment. To date, two main strategies have been employed for generating these structures: (1) self-assembly, in which similar atomic arrangements on the surface of primary nanoparticles approach each other, and then fuse together by oriented attachment; (2) anisotropic growth, in which different capping agents, surfactants, or templates are used to induce an anisotropic growth on the surface of the available seeds or newly generated particles. Both these strategies require multi-stage, complex, and time-consuming processes, using hazardous chemical reagents. In our previous research we demonstrated the creation of hybrid multi-yolk-shell Au-Ag@a-C:H nanoparticles, found to be efficient for detecting ultra-low-volumes and concentrations of biological and hazardous impurities [16, 17]. The nanoparticles were prepared as a result of laser-induced transformation of supramolecular complexes - being precursor and building blocks for the formation of complex nanohybrids.

In the present communication we report on the direct one-step synthetic method for generating hybrid Au-Ag/C nanostructures of the desired morphology. The method provides in situ planting of such hybrid structures as nanoparticles (NPs), nanoflakes (NFKs) and nanoflowers (NFWs) on crystalline substrates. This method is efficient and time-saving, does not require application of external electric or magnetic fields, nor a thermal treatment. The method is based on laser irradiation of the substrate/solution interface resulting in the formation of hybrid nanostructures of different morphology in the laser-treated area of the substrate. Easily controllable technological parameters of the process (composition and concentration of the liquid phase, crystalline vs. amorphous substrates and laser parameters such as irradiation intensity and dwell time) allow for targeted formation of the hybrid Au-Ag/C NPs, NFKs and NFWs. The scheme of the deposition process is shown in Fig. 1. Laser radiation is directed to the substrate-solution interface from the side of the solution (Fig. 1a). In case of a transparent substrate it is also possible to illuminate through the substrate (Fig. 1b). As a liquid phase we used solutions of a supramolecular complex $[\{Au_{10}Ag_{12}(C_2Ph)_{20}\}Au_3(PPh_2(C_6H_4)_3PPh_2)_3][PF_6]_5$ (hereinafter termed SMC) in different solvents (acetone, dichloroethane and acetofenone, all of analytical grade purity The supramolecular complex (Fig.S1a) consists of a central heterometallic cluster core (ca. 2 nm) surrounded by phosphine and alkyl ligands; it was synthesized according to the published procedure [18].

Laser wavelength for the deposition process is chosen in accordance with the characteristic absorption bands of SMC (Fig. S1b) corresponding with (i) intraligand (IL) electron transition (200-300 nm), (ii) metal-to-ligand charge transfer (MLCT) (300-350 nm), and (iii) transitions between orbitals of the cluster core (MMCT) (370-450 nm). All these bands can be used for deposition process, however IL band (200-300 nm) is typically not usable due to high absorbance of majority of the system components and materials (solvents, reaction vessels, etc.) in this spectral region. MMCT band has a rather low absorption cross-section that results in low efficiency of energy absorption. This is why the He-Cd laser radiation (325 nm wavelength) was used for optical excitation of SMC into the 300-350 nm MLCT absorption band. A low intensity (0.1 mW/cm$^2$) CW laser can be used, that provides mild illumination conditions and makes possible to avoid side thermal effects (delocalization of the deposition process, thermal decomposition of the deposits, etc.). The temperature in the laser-affected area was measured with Thermovision camera Ti32 (Fluke) and found to be 26˚C. The deposit formation occurs in the laser-affected area of the substrate as a result of laser irradiation for 5 – 20 min. It was found that the deposition process can be realized for all the used solvents (acetone, acetophenone, dichloroethane) in a wide concentration range (2 – 10 mg/ml). Data given in Table S1 indicate that nanoparticles are the most easily planted structures; their deposition takes place in all systems studied except for low concentrations of dichloroethane and high concentrations of acetophenone solutions.

Along with NPs' formation, deposition of nanoflakes was also observed for all acetophenone solutions and highly concentrated dichloroethane solutions. Flowers were formed in acetophenone at 4 and 5 mg/ml concentrations and at 10 mg/ml in dichloroethane. It is worth noting that glass cover slides with a conductive indium-tin-oxide (ITO) film were used as the substrate for laser-induced deposition of NPs, NFWs and NFKs; while uncoated glass cover slides could only host NPs formation. The typical images of the obtained structures are presented in Fig. 2.

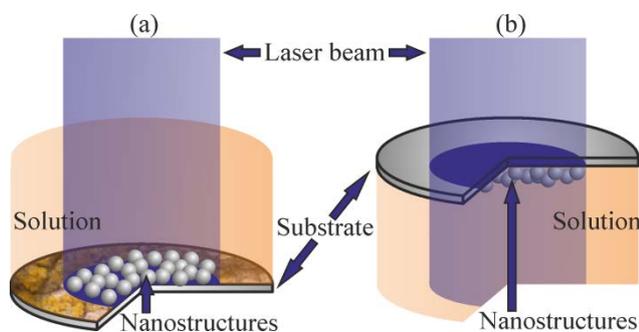

**Figure 1** The scheme of laser-induced deposition (a) the scheme with nontransparent substrate for laser radiation, (b) the scheme with transparent substrate for laser radiation.

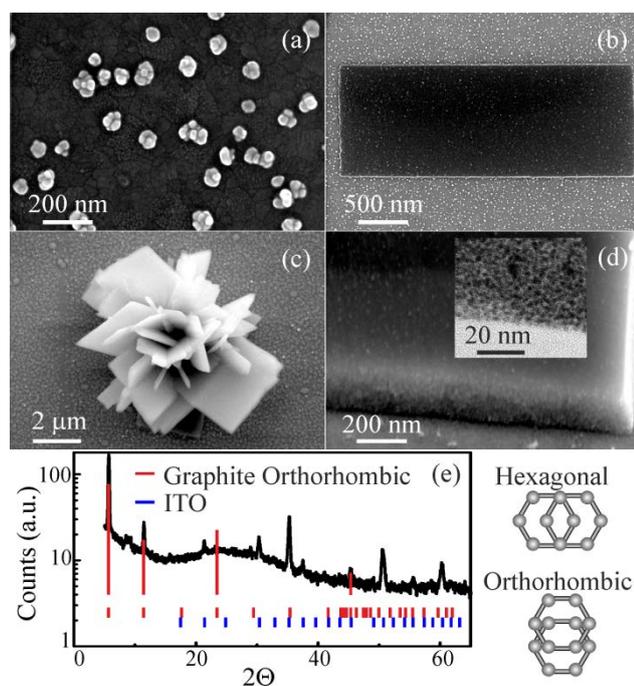

**Figure 2** SEM images of (a) nanoparticles, (b) nanoflake, and (c) nanoflower obtained at the surface of ITO-covered glass as a result of laser-induced deposition from solutions of supramolecular complexes. (d) SEM image of a nanoflake, the insert is a TEM image where many metal nanoclusters show up as dark spots in the crystalline carbon matrix, (e) XRD of the sample with deposited nanoflakes, nanoflowers and line diagrams of orthorhombic graphite (red) and ITO (blue), insert shows schematic presentation of hexagonal and orthorhombic graphite.

The deposited NPs were similar in all cases (Fig. 2a); the size of NPs was found to be 20-60 nm. In accordance with our previous investigations, the NPs formed have complex morphology: they consist of Au−Ag nanoalloy clusters of 3-5 nm in diameter embedded in a matrix of hydrogenated amorphous carbon a-C:H (Au-Ag@a-C:H NPs) [16, 17].

Nanoflakes are atomically smooth regularly shaped flat structures of 1-2 μm times 3-4 μm size with a thickness of 10 – 100 nm as determined from high resolution scanning electron micrographs (SEM) (Fig. 2b). Nanoflowers are composed of nanoflakes arranged like a flower standing on their thin edges (cf. Fig. 2c).

The EDX analysis demonstrates that all the deposited structures (NPs, NFWs and NFKs) consist of carbon, gold and silver with a typical component ratio 90/5/5, coinciding with the stoichiometry of the SMC precursor (92/4/4) (Fig. S2). Normalized absorption spectra of obtained nanostructures are shown in Fig. S3. The position of the absorption band is typical for the Au−Ag surface plasmon resonance (SPR) and is indicative of the alloy composition of the metallic phases formed in the NPs [19]. SEM and transmission electron microscopy (TEM) data presented in Fig. 2d demonstrate that in analogy to NPs, nanoflakes also consist of metal nanoclusters stochastically distributed in a carbonaceous matrix. The observed absorption was polarization independent indicating the spherical geometry of the metal nanophase. But in case of nanoflakes and nanoflowers, carbon was found to be crystalline. The x-ray diffraction (XRD) pattern of the samples with deposited NFWs and NFKs (Fig. 2e) is close to the pattern of orthorhombic graphite (ICSD Collection Code: 28419), the only misfit was found for reflection (008) 2Θ = 24.5712° that can be related to the nanoscale thickness of the studied objects for the [001] direction. Orthorhombic graphite is considered to be a non-centrosymmetric crystalline phase with optical activity; preliminary measurements confirm this expectation. Orthorhombic graphite is a rare species of graphite that differs in symmetry from the hexagonal allotrope (which is centrosymmetric) in the relative displacement of neighboring graphite planes (see insert in Fig. 2e). Orthorhombic graphite is considered to be an intermediate phase forming at extreme conditions in the course of graphite-to-diamond transition. The only alternative approach reported in literature so far allows for the creation of orthorhombic graphite bilayer structures only. It is based on the unzipping of carbon nanotubes via a chemical multistep treatment procedure [20]. Nonetheless, no approach for the fabrication orthorhombic graphite with metal inclusions has been reported to date. As graphite is a layered structure, the metal Au-Ag nanoclusters most likely occupy interlayer positions.

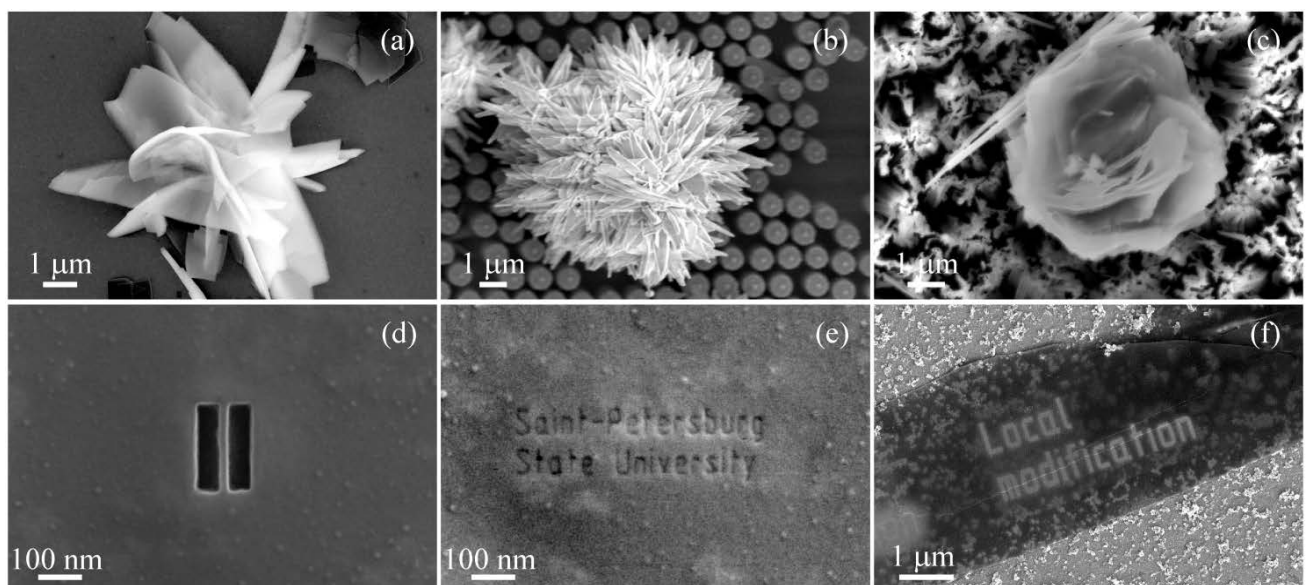

**Figure 3** SEM images of nanoflowers obtained by growth on (a) the surface of ITO-covered glass, (b), (c) different silicon nanowires with diameters between 800nm and 10nm; Helium ion microscope images of modified nanoflakes: (d), (e) flakes cut with focused ion beams, (f) local modification of a nanoflake with focused helium-ion beam.

Raman spectroscopy of the planted hybrid Au-Ag/C nanostructures revealed features typical for carbons – G and D peaks, which lie around 1560 and 1360 cm$^{-1}$, respectively. Fig.S4 presents Raman spectra of the obtained nanoparticles and flakes. Characteristic G peaks for the NPs and NFKs coincide precisely while spectral positions and intensities of D peaks indicate that the nanoparticles have higher structural disorder than flakes. Taking the crystalline structure of the carbon phase in the Au-Ag/C flakes into account one can expect a Raman spectrum that is characterized by low intensity D peaks and the associated low $I_D/I_G$ value [21]. However, the D peak originates from breathing modes of six-atom rings and requires defects for its activation. In our case the D peak in the Raman spectrum of the NFKs is likely activated by Au-Ag nanoclusters.

It is worthwhile to note that the morphology of the flowers can be controlled by the substrate used for deposition. The laser-induced planting process was realized in three different ways with ITO coated microscope cover glass and with structured monocrystalline Si nanowires of two very different diameters, ~800nm and ~ 10-30 nm [22]. Fig. 3 shows the variation of flower morphology with substrate properties: the structure of the different types of nanowires leads to local electric fields dictating the shape of flowers growing on the respective substrate surface. This allows for the formation of spiky petal structure (chrysanthemum-shaped) in the case of the more regular Si nanowire structures (Fig. 3b) and of rose-shaped flowers for the more chaotic Si nanowire geometry (Fig. 3c). The planted multi-petal structures with complex shapes are anisotropic hybrid metal/carbon nanostructured materials that are expected to provide numerous extraordinary properties not possible for spherical analogues [23].

The interest in and the demand for high quality flake-like nanostructures as basic building blocks is fueled by the development of optical nano-circuitry which makes use of the plasmonic resonances of metal nanostructures and the associated strongly enhanced local fields [24–26]. Fig. 3(d-f) give Helium ion microscope (HIM) images of the flakes designed with focused ion beam. The planted hybrid Au-Ag@C flakes can be cut by means of sputtering with a focused helium ion beam, and a resolution of 18 nm can be achieved for 15 nm-thick flakes (Fig. 3d) thus demonstrating a new material suited for fabricating subwavelength devices. An example of more elaborate cutting is shown in Fig. 3e. It is noteworthy that the flakes can be patterned not only by cutting but also by locally modifying the material with the helium ion beam using a fluence of about $10^{16}$ cm$^{-2}$ which is several orders of magnitude lower than the ion fluence required for sputtering ($10^{19}$ cm$^{-2}$). The shades of grey in Fig.3f reflect the extent to which the local secondary electron yield is enhanced, which can be attributed to a local modification of the material conductivity and/or material work function implying a local modification of the electronic structure of the flake. Thus, complex planar elements can be created without cutting the material by means of local irradiation of flakes with the focused helium ion beam. The combination of well-designed architecture and locally modified properties of optical nano-circuitry elements with plasmonic bimetal nano inclusions forming controllable hot-spot areas offers new opportunities when designing high-density optical devices involving e.g. plasmon enhanced transmission through sub wavelength holes [27] or selectively exciting sub-wavelength magnetic dipoles [28].

Based on the demonstrations described above we propose developing a controlled direct laser planting process for forming hybrid carbon/metal nanostructures of desired morphology (nanoparticles, nanoflakes, nanoflowers). The planting process can easily be controlled by deposition parameters such as composition and concentration of liquid phase, nature of substrate (crystalline/amorphous), irradiation dwell time and deposition geometry. The attractiveness of the suggested synthetic methodology consists in requiring only a single step for growing hybrid nanostructures, which are complex in composition (Au-Ag nanoclusters in amorphous or crystalline carbonaceous matrix) and morphology (2D atomically smooth flakes or 3D morphologically complex multi-petal hybrid nanostructures). A distinguishing feature of the suggested approach is the open-air and room-temperature fabrication process resulting from photo-induced transformation of the SMC precursor. It should be noted that the mechanism of formation of the crystalline carbonaceous structures with incorporated bimetallic nanoclusters is still unknown and challenging. We will further study the dynamics of this process in more detail trying to better understand the laser planting process the feasibility of which we already demonstrated.

**Acknowledgement**

The reported study was supported by Ministry of education and science of Russia within project 14.604.21.0078 (identification number RFMEFI60414X0078). Raman, and UV/VIS absorption spectra were measured at Center for Optical and Laser Materials Research, TEM and SEM analysis was carried out at the Interdisciplinary Resource Center for Nanotechnology, XRD analysis was carried out at Centre for X-ray Diffraction Studies, St. Petersburg State University.

**Electronic supplementary material:** Supplementary material (Scheme of laser-induced deposition process; SMC complex, SMC absorption spectrum; EDX, absorption, Raman spectra of nanostructures; Table of liquid phase composition; Nanohybrid characterization) is available in the online version of this article at http://dx.doi.org/10.1007/s12274

**Electronic Supplementary Material**

**Direct laser planting of hybrid Au-Ag/C nanostructures - nanoparticles, flakes and flowers**

A. Manshina[1](✉), A. Povolotckaia[2], M. Bashouti[3], A. Povolotskiy[1], Yu. Petrov[4], I. Koshevoy[1,5], S. Christiansen[3,6], S. Tunik[1], and G. Leuchs[3,7,8]

[1] Institute of Chemistry, St. Petersburg State University, Universitetskii pr. 26, St. Petersburg, 198504, Russia
[2] Center for optical and laser materials research, Research park, St. Petersburg State University, Ulianovskaya St. 5, St. Petersburg, 198504, Russia
[3] Max Planck Institute for the Science of Light, Günther Scharowsky-Str. 1 Bau 24, D - 91058 Erlangen, Germany
[4] Interdisciplinary Resource Center for Nanotechnology, Research park, St. Petersburg State University, Ulianovskaya St. 1, St. Petersburg, 198504, Russia
[5] University of Eastern Finland, Joensuu, 80101, Finland
[6] Helmholtz Center Berlin for Materials and Energy, Hahn-Meitner-Platz 1, 14109 Berlin, Germany
[7] Department of Physics, Friedrich-Alexander-Universität Erlangen-Nürnberg, D-91058 Erlangen, Germany
[8] Department of Physics, University of Ottawa, 140 Louis Pasteur (130), Ottawa, ON, K1N 6N5, Canada

**Experimental section**
**Laser deposition of hybrid Au-Ag/C nanostructures**

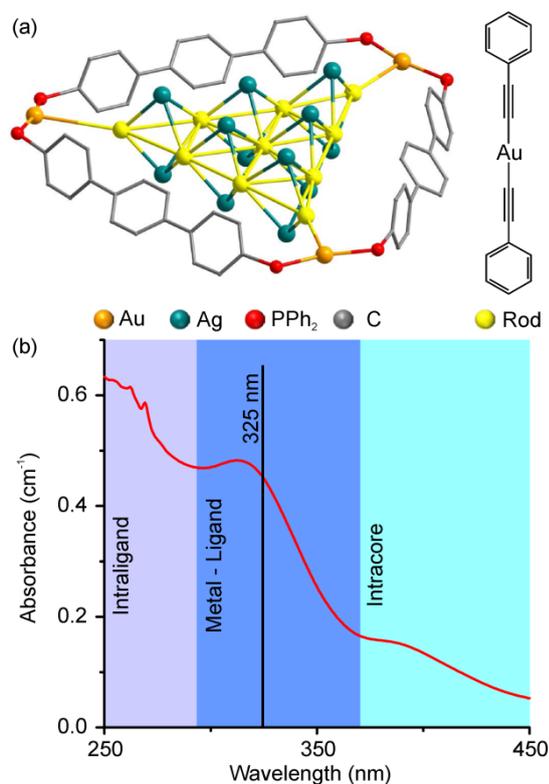

**Figure S1** (a) The scheme of supramolecular complex
$[\{Au_{10}Ag_{12}(C_2Ph)_{20}\}Au_3(PPh_2(C_6H_4)_3PPh_2)_3][PF_6]_5$.
(b) Absorption spectrum of supramolecular complex in dichloroethane solution.

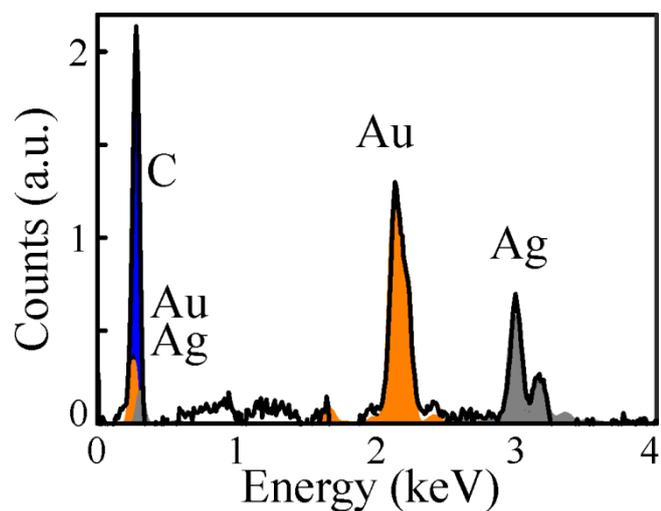
**Figure S2** EDX analysis of deposited nanostructures.

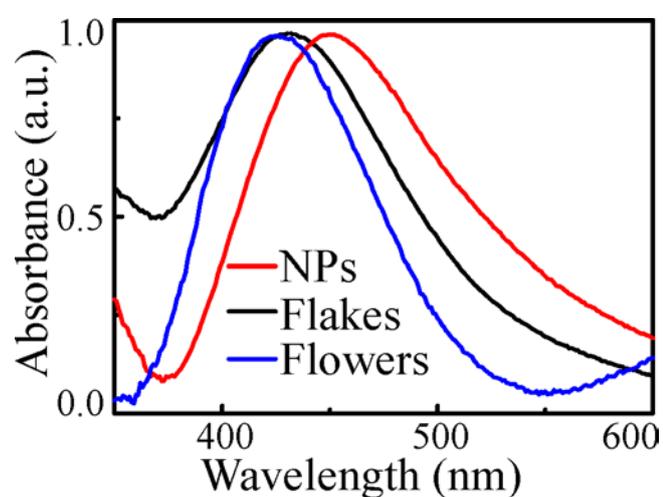
**Figure S3** Normalized absorption spectra of obtained nanostructures.

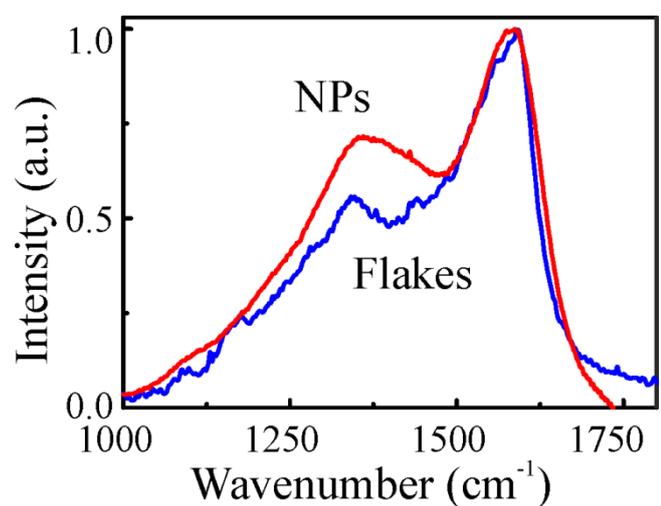
**Figure S4** Raman spectra of several nanoparticles (red line) illuminated simultaneously and of a single flake (blue line).

**Table S1** Composition of the liquid phase and morphology of the planted hybrid nanostructures

| SMC, mg/ml | Acetone | Aceto-phenone | Dichloroethane |
|---|---|---|---|
| 2 | NPs | NPs Flakes | - |
| 3 | NPs | NPs Flakes | - |
| 4 | NPs | NPs Flakes Flowers | NPs |
| 5 | Agglom. NPs | NPs Flakes Flowers | NPs |
| 10 | Agglom. NPs | - | Agglom. NPs Flakes Flowers |

**Nanohybrid characterization**

A scanning electron microscope Zeiss Merlin was used for all SEM based characterization with 10 kV acceleration voltage used for imaging and EDX analysis. A transmission electron microscope Zeiss Libra 200FE was used at 200 kV for all TEM measurements. Local modification and cutting of the flakes were performed with the helium ion microscope Zeiss Orion equipped with the Nanomaker pattern generator. Measurements of the absorption spectra of planted nanostructures were performed using a double-beam spectrophotometer Perkin Elmer Lambda 1050 using an Ulbricht sphere. The Raman measurements were performed using a Raman spectrometer T64000 (Horiba) equipped with confocal microscope. The Raman spectra were recorded with spatial resolution using solid state laser $\lambda_{ex}$ = 532 nm, power 0.1 mW at temperature 10 K. The laser was focused onto an approximately 2 µm-diameter spot at the sample surface. Raman spectra of NPs and NFKs were measured separately. The spectra were collected over 1 hour in back reflection geometry through a 40X0.5NA objective, the reflected signal was detected with thermo-electrically cooled CCD array. X-ray phase analysis of the samples was carried out using HR diffractometer D8 DISCOVER (Bruker) with CuKa1 radiation ($\lambda_{CuK\alpha1}$ = 1.54056 Å). Measurements were performed in the range of 2Θ = 5–70˚. Data processing and phase identification was carried out in the program PDXL2 (Rigaku Corporation) using a powder diffraction database PowderDiffractionFile (PDF-2, 2011)